%% file: paper.tex
\documentclass[prb, aps,reprint,superscriptaddress,titlepage,floatfix]{revtex4-1}
\usepackage{graphicx}
\usepackage{nicefrac}
\usepackage{amsfonts}
\usepackage[version=3]{mhchem}
\usepackage{amssymb}
\usepackage{amsmath} 
\usepackage{subfigure}
\usepackage{multirow} 
\usepackage{tabularx} 
\usepackage{array}
\usepackage{units}
\usepackage{braket}
\usepackage{bm,times}
\usepackage{booktabs}
\usepackage{enumitem}
\usepackage{mathtools}

\usepackage[colorlinks = true, linkcolor = blue, urlcolor  = blue, citecolor = blue, anchorcolor = blue]{hyperref}

\def\be{\begin{equation}}
\def\ee{\end{equation}}
\def\bea{\begin{eqnarray}}
\def\eea{\end{eqnarray}}

\def\degree{$^\circ$}
\def\etal{\it et al.\,}
\def\asi{{\it a}-Si}
\def\csi{{\it c}-Si}

\begin{document}

\title{
Atomistic simulation of nearly defect-free models of amorphous 
silicon: An information-based approach
}

\author{Dil K. Limbu}
\email{dil.limbu@usm.edu}
\affiliation{
Department of Physics and Astronomy, The University of Southern 
Mississippi, Hattiesburg, Mississippi 39406
}

\author{Raymond Atta-Fynn}
\email{attafynn@uta.edu}
\affiliation{Department of Physics, University of Texas at Arlington, Arlington, Texas 76019}

\author{Parthapratim Biswas}
\email[Corresponding author: ]{partha.biswas@usm.edu}
\affiliation{Department of Physics and Astronomy, The University of Southern
Mississippi, Hattiesburg, Mississippi 39406}

\begin{abstract}
We present an information-based total-energy optimization method to
produce nearly defect-free structural models of amorphous silicon.
Using geometrical, structural and topological information from
disordered tetrahedral networks, we have shown that it is possible to
generate structural configurations of amorphous silicon, which are
superior than the models obtained from conventional reverse Monte
Carlo and molecular-dynamics simulations. The new data-driven hybrid
approach presented here is capable of producing atomistic models with
structural and electronic properties which are on a par with those
obtained from the modified Wooten-Winer-Weaire (WWW) models of
amorphous silicon. Structural, electronic and thermodynamic
properties of the hybrid models are compared with the best dynamical
models obtained from using machine-intelligence-based algorithms and
efficient classical molecular-dynamics simulations, reported in the
recent literature. We have shown that, together with the WWW models,
our hybrid models represent one of the best structural models so far
produced by total-energy-based Monte Carlo methods in conjunction with
experimental diffraction data and a few structural constraints.  
\end{abstract}
\maketitle 

\section{Introduction}

Amorphous silicon continues to play a major role in the application of
silicon-based device technology.~\cite{street2000} Recent developments
of silicon-based heterojunction intrinsic technology
(HIT)~\cite{taguchi2014} for photovoltaic cells and the two-qubit
quantum logic gates~\cite{veldhorst2015} are indicative of the
continuing importance of amorphous/crystalline silicon and
silicon-based materials. While the structure of amorphous silicon
({\asi}) can be readily described by using a continuous 
random network (CRN) model,\cite{zachariasen1932} which is characterized by the presence
of high degree of tetrahedral ordering, the construction of 
CRN models with minimal strain and few coordination defects has been a
vexing problem in structural modeling of amorphous silicon. Until
recently, high-quality structural models of {\asi} are best produced
by the Monte Carlo based bond-switching algorithm of Wooten, Winer 
and Weaire (WWW).~\cite{WWW1985} The subsequent modification of the
algorithm by Barkema and Mousseau~\cite{barkema2000} further augments
the capability of the WWW method in producing large structural models
of {\asi}, exhibiting experimentally compliant structural, electronic 
and vibrational properties. Other important approaches that are often
employed to simulate {\asi} are molecular-dynamics (MD)
simulations~\cite{deringer2018,atta-fynn2018,pedersen2017,car1988}
(either ab initio/classical MD or an intermediate approach between the
two), reverse Monte Carlo simulations
(RMC),~\cite{biswas2004,RMC2001,gereben1994} and the recently
developed hybrid RMC methods in various
flavors~\cite{cliffe2010,pandey2015,biswas2005} including the Force
Enhanced Atomic Refinement (FEAR) approach.~\cite{pandey2015} In the
following, we present briefly an information-driven inverse approach
(INDIA),~\cite{limbu2018} where we have shown that high-quality
structural models of amorphous silicon can be produced efficiently by
inverting experimental scattering data, such as structure-factor 
or pair-correlation data, along with additional geometrical and 
topological information on the local bonding environment of silicon 
atoms in the amorphous state. 

The rest of the paper is as follows. In Sect.\,II, we briefly 
outlined the computational method for structural modeling of 
amorphous silicon using classical and quantum-mechanical 
simulations. Section III addressed the results and discussion 
with an emphasis on the structural, electronic, vibrational 
and the thermodynamic properties of the models. This is followed 
by the conclusions of our work. 
 
\section{Computational Method}

Recently, we have implemented an information-driven inverse approach
(INDIA) to model the atomistic structure of amorphous silicon. A detailed
description of the method can be found in Ref.\,\onlinecite{limbu2018}.  
Here, we primarily discussed the characteristic features of the 
resulting models with particular emphasis on the structural, 
topological, vibrational and thermodynamic properties of the models. 
Hereafter, we refer to these models as 
INDIA models. Following Ref.\,\onlinecite{limbu2018}, it can be shown 
that high-quality structural models of {\asi} can be generated by 
inverting experimental diffraction data and structural constraints in
conjunction with an appropriate total-energy functional. For amorphous 
silicon, the method can be illustrated by employing the modified
Stillinger-Weber (SW) potential,~\cite{stillinger1985,vink2001}
which is given by,
\be
\label{sw}
V(\mathbf R^N) = \frac{1}{2}\sum_{i=1}^N\sum_{\substack{j=1\\(j\neq i)}}^N v_2(r_{ij})
+ \sum_{i=1}^N \sum_{\substack{j=1\\(j\neq i)}}^N 
\sum_{\substack{k=1\\(k\neq i)\\(k > j)}}^N v_3(\bm{r}_{ij},\bm{r}_{ik}). 
\ee
\noindent 
In Eq.\,\ref{sw}, the two-body contribution, $ v_2 (r_{ij})$, is given by, 
\be
    v_2(r_{ij})=
    \epsilon A\left[B\left(\frac{r_{ij}}{\sigma}\right)^{-p}-1\right]
\exp{\left(\frac{\sigma}{r_{ij}-a\sigma}\right)} \Theta(a\sigma - r_{ij}),
\ee

\noindent 
and the three-body contribution follows from, 
\be
\begin{split}
    v_3(\bm{r}_{ij},\bm{r}_{ik})&=
   \epsilon\lambda\left(\cos\theta_{jik} + \frac{1}{3}\right)^2
  \exp{\left(\frac{\sigma\gamma}{r_{ij}-a\sigma}
+ \frac{\sigma\gamma}{r_{ik}-a\sigma}\right)} \\ 
  &\times \Theta(a\sigma - r_{ij})\Theta(a\sigma-r_{ik}),
\end{split}
\ee
\noindent 
with $\Theta$ being the Heaviside step function. The potential
parameters, due to Vink {\etal},~\cite{vink2001} were used in this
work. The simulations were carried out by maintaining a minimum
distance of 2.0 {\AA} between silicon atoms and the mass density
of the model was matched with the experimental density of 2.25 g/cm$^3$ 
for amorphous silicon. 
The electronic and vibrational properties of models were
studied using the local-basis density-functional theory (DFT) code
{\sc Siesta}.~\cite{siesta2002} The latter employs pseudoatomic
orbitals as basis functions (employing the double-zeta basis functions 
in the present work) and the norm-conserving Troullier-Martins 
pseudopotentials~\cite{troullier1991} within the
Perdew-Burke-Ernzerhof (PBE) formulation~\cite{perdew1996} of the
generalized gradient approximation (GGA).

\section{Results and Discussion}

In this section, we discussed the results from the INDIA models 
and compared with the corresponding WWW
models. In addition, we also examined the results from 
molecular-dynamical models obtained using machine-intelligence-based 
algorithms (ML-MD)~\cite{deringer2018} and molecular-dynamics 
simulations (SW-MD).~\cite{atta-fynn2018} We 
studied the structural properties of {\asi} models consisting of up to
1024 atoms and presented the results in Table~\ref{TAB1}, which listed
various structural properties of the relaxation-based INDIA models, 
WWW models and MD-based SW models showing the average bond length
({\textless}$r${\textgreater}), average bond angle
({\textless}$\theta${\textgreater}) and the percentage of 4-fold
coordination number (C$_4$), as well as the effective coordination
number (ECN). The last two quantities were computed by using the 
relations described in Refs.\,\onlinecite{hoppe1979,limbu2018a}.

\begin{table}[t!]
\caption{\label{TAB1} 
Structural properties of INDIA, SW-MD and WWW models: average bond
length {\textless}$r${\textgreater}, average bond angle
{\textless}$\theta${\textgreater}, the root-mean-square (RMS) 
deviation of the bond angles $\Delta\theta$, the percentage 
of 4-fold coordination number C$_4$ and the effective 
coordination number (ECN), respectively.}
\begin{ruledtabular}
\begin{tabular}{lccccccc}
Model &N &{\textless}$r${\textgreater} &$ \langle\theta\rangle$ &$\Delta\theta$ &$(\Delta\theta_G)$\footnote{Values obtained from the Gaussian approximation of the
bond-angle distribution.} & C$_4(\%)$  &ECN \\
\hline
INDIA &\multirow{3}{*}{300} &2.386 &109.10 & 11.41 &10.42  &99.33 &3.956\\
SW-MD & &2.380 &109.22 & 9.31 &8.44   &99.33  &3.973\\
WWW & &2.378 &109.18 & 10.44 &10.13  & 100.0  &3.963\\ \addlinespace[0.5 mm]\hline
INDIA &\multirow{4}{*}{512} &2.387 &109.10 & 11.48 &10.64  &99.60  &3.955\\
SW-MD & &2.379 &109.27 & 9.12 &8.58 &99.22  &3.976\\
ML-MD\footnote{From Ref.\,\onlinecite{deringer2018}} & &2.371 &109.19 & 9.69 &9.36  &98.44  &3.940\\
WWW & &2.365 &109.11 & 10.69 &10.47  &100.0 &3.974\\ \addlinespace[1mm]\hline
INDIA &\multirow{3}{*}{1024} &2.390 &109.01 & 11.96 &10.78 &98.34  &3.958 \\
SW-MD & &2.381 &109.27 & 8.94 &8.46 & 99.22  &3.968\\
WWW & &2.371 &109.14 & 10.63 &10.30 &100.0  &3.969\\
\end{tabular}
\end{ruledtabular}
\end{table}

\begin{figure}[t!]
\includegraphics[width=0.45\textwidth]{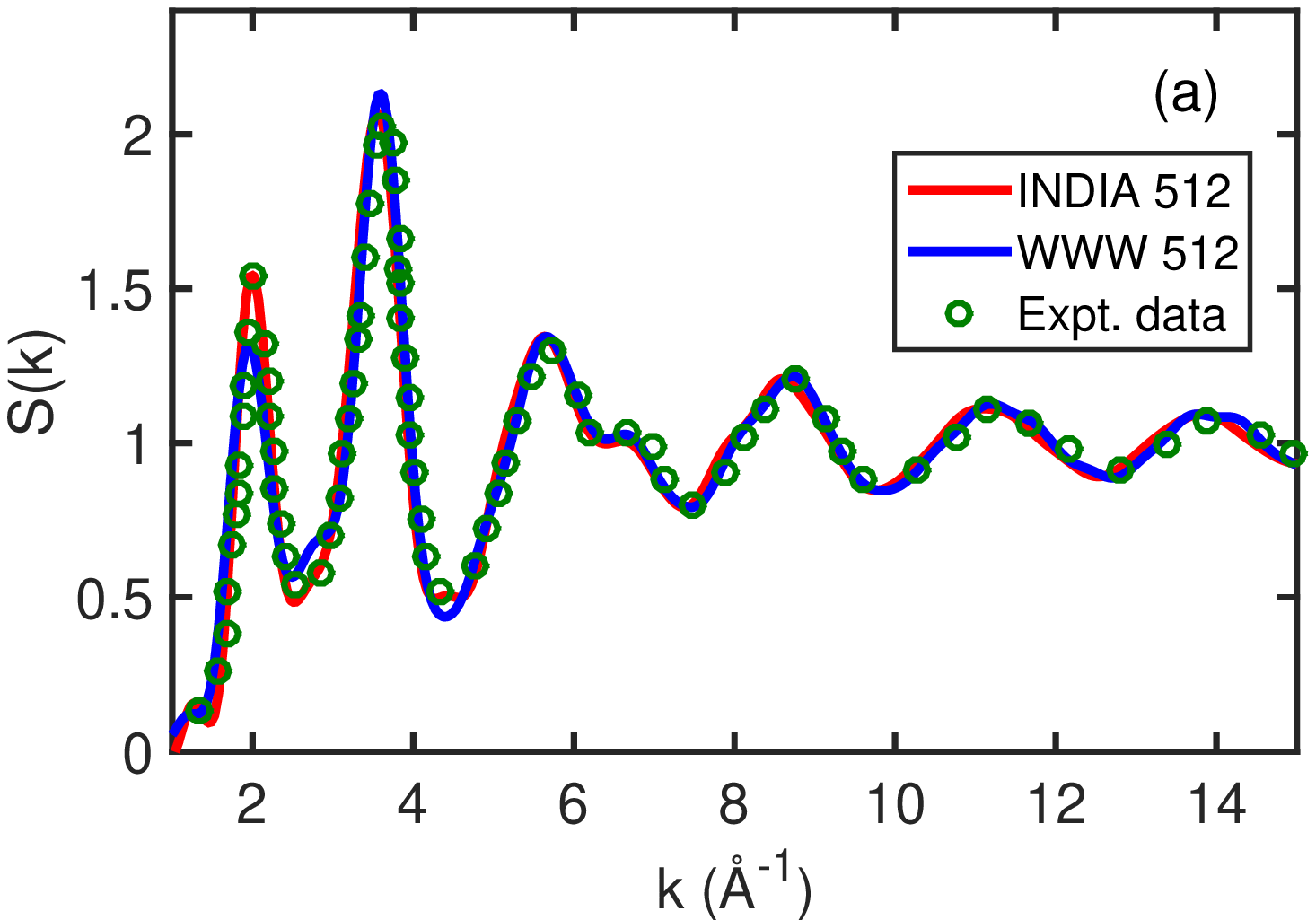}\vspace{2mm}
\includegraphics[width=0.45\textwidth]{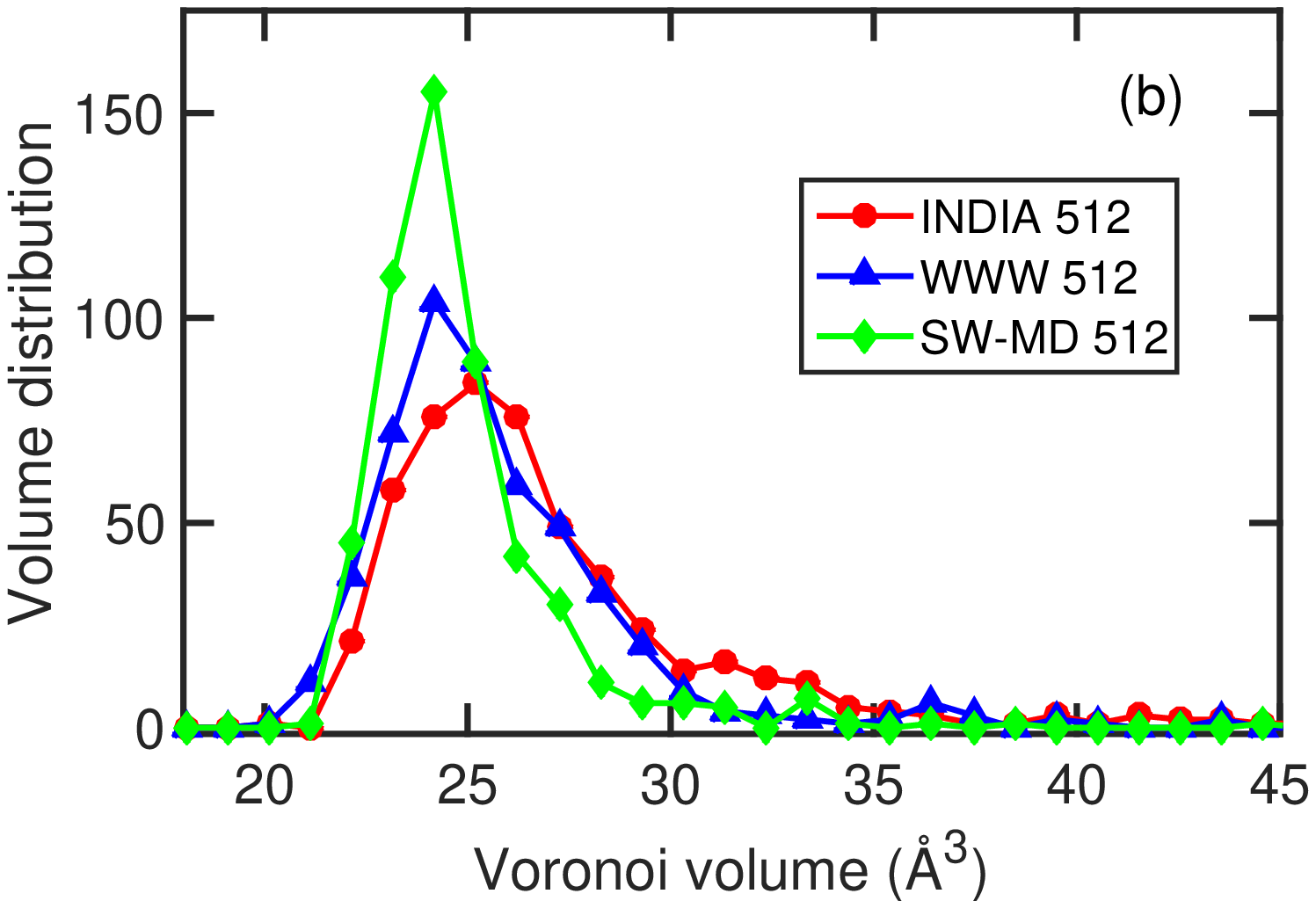}
\caption{\label{fig1}
{\small 
(a) Structure factor, S($k$), from the 512-atom INDIA (red) and 
WWW (blue) models compared with the experimental data for as-deposited 
{\asi} samples from Laaziri {\etal},~\cite{laaziri1999} 
reprinted with permission from American Physical Society. 
(b) The (approximate) distributions of Voronoi volumes 
obtained from the INDIA (red), WWW (blue) and SW-MD (green) models. 
For references to color in this figure legend, see the 
web version of this article. 
}}
\end{figure}

The static structure factor, S($k$), of the INDIA models was 
compared with the WWW models and the experimental 
structural-factor data from Laaziri {\etal}~\cite{laaziri1999} 
The structure factors presented in Fig.\,\ref{fig1}(a) show that the 
results from the 512-atom INDIA model agree with the 
WWW model of identical size, as far as the two-body correlations 
are concerned. 
The results also match with the experimental 
data for as-deposited {\asi} samples from 
Laaziri {\etal}~\cite{laaziri1999} To examine the 
interstitial volumes associated with 
the model networks,  we computed the Voronoi volumes 
(of the atoms) of the 512-atom INDIA, WWW and SW-MD 
models and plotted the respective Voronoi-volume distributions 
in Fig.\,\ref{fig1}(b). The following observations are 
now in order. First, the Voronoi-volume distribution 
of a model can provide additional information on 
the degree of local ordering of a disordered 
network in a manner similar to the pair-correlation 
and bond-angle distributions. A narrow distribution 
with a sharp peak is indicative of more ordered 
networks, whereas networks with less order tend to show 
a relatively broad distribution. 
Second, since the Voronoi volume associated with 
an atom depends on the spatial positions of its 
nearest neighbors,  the degree of inhomogeneity 
of a disordered network can be approximately gauged 
by analyzing the volume distribution, particularly 
the region away from the central peak.  A large 
(or small) Voronoi volume is indicative of the presence 
of sparse (or dense) atomic environment in the 
network.~\cite{note1} Thus, the presence of microvoids 
and the density-deficient regions in 
the network can be readily manifested in the 
Voronoi-volume distribution. It is apparent from 
Fig.\,\ref{fig1}(b) that, as far as the Voronoi 
volumes or the interstitial regions of the atoms are 
concerned, the SW-MD models exhibit more order than 
the corresponding WWW and INDIA models. This observation 
is quite consistent with the fact that MD models 
generally represent annealed samples of {\asi} more 
closely, and hence more ordered, than those obtained 
from the relaxation-based Monte Carlo or similar approaches. 

Since the bond angles between the nearest-neighbor atoms provide 
limited three-body correlations between atoms, we analyze the 
bond-angle distribution (BAD) of the atoms in the networks. The 
distribution is found to be essentially Gaussian in
nature, which gives the average bond angle close to the value 
of ideally tetrahedrally bonded silicon atoms of 109.47{\degree} in
crystalline silicon and the root-mean-square (RMS) deviation of the
bond angles ($\Delta\theta$) in the experimentally observed range of
9-11{\degree}.~\cite{beeman1985} To obtain $\Delta\theta$, we fitted
the BAD with a Gaussian distribution. Although the Gaussian
approximation to the bond-angle distribution somewhat
underestimates the RMS value, due to the absence of a few large/small
angles in the distribution, it describes the BAD fairly accurately for
high-quality {\asi} networks. This is reflected in Table~\ref{TAB1}.

\begin{figure}[t!]
\includegraphics[width=0.45\textwidth]{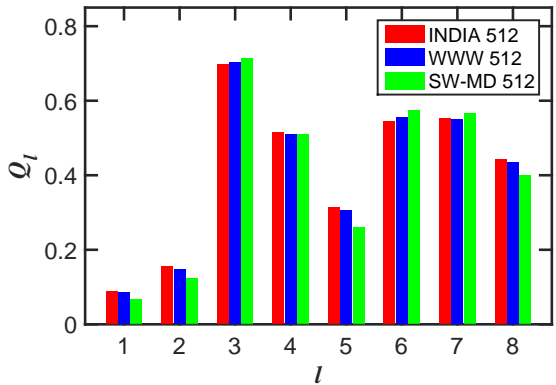}
\caption{\label{fig2}
{\small 
The distribution of the bond-orientational order parameter, $Q_l$, 
for the 512-atom INDIA (red), WWW (blue)  and SW-MD (green) models. 
For references to color in this figure legend, see the web version 
of this article.
}}
\end{figure}
Further characterization of the network is possible by examining the
bond-orientational order parameter (BOP), $Q_l$, as defined by
Steinhardt {\etal},~\cite{steinhardt1983} which provides information
on the orientation of a group of bonds. Figure~\ref{fig2} shows the
$Q_l$ values, for $l$ = 1 to 8, for INDIA, WWW and SW-MD models
consisting of 512 atoms. The BOP not only incorporates some aspects of
structural information from higher-order correlation functions, but
also provides a simple and effective measure for determining the
presence of microcrystalline or paracrystalline structural units in
the networks.~\cite{limbu2017b} Here, we have used  $Q_l$ ($l$=1,2,5) to
determine the {\textit{degree of crystallinity}. While the magnitude
of  $Q_1$, $Q_2$ and  $Q_5$ are exactly zero for ideal
{\csi}~\cite{atta-fynn2018} networks, a high value of  $Q_5$ for the INDIA and
WWW models indicates more amorphous or disordered nature of these
models in comparison to the SW-MD model. This observation is in 
agreement with the conclusion followed from the Voronoi-volume 
distribution as shown in Fig.\,\ref{fig1}(b). 

\begin{figure}[t!]
\includegraphics[width=0.45\textwidth]{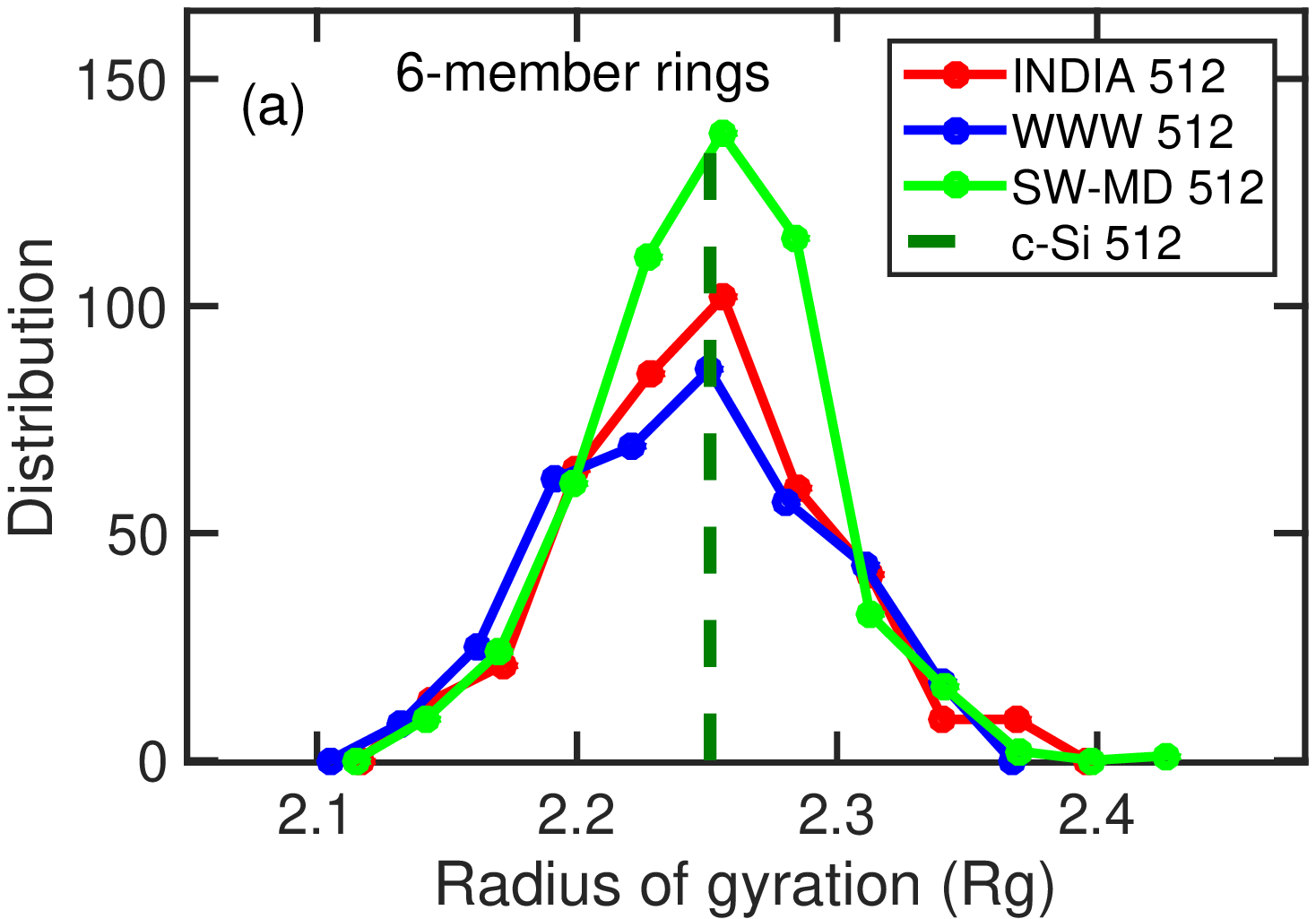}\vspace{2mm}
\includegraphics[width=0.45\textwidth]{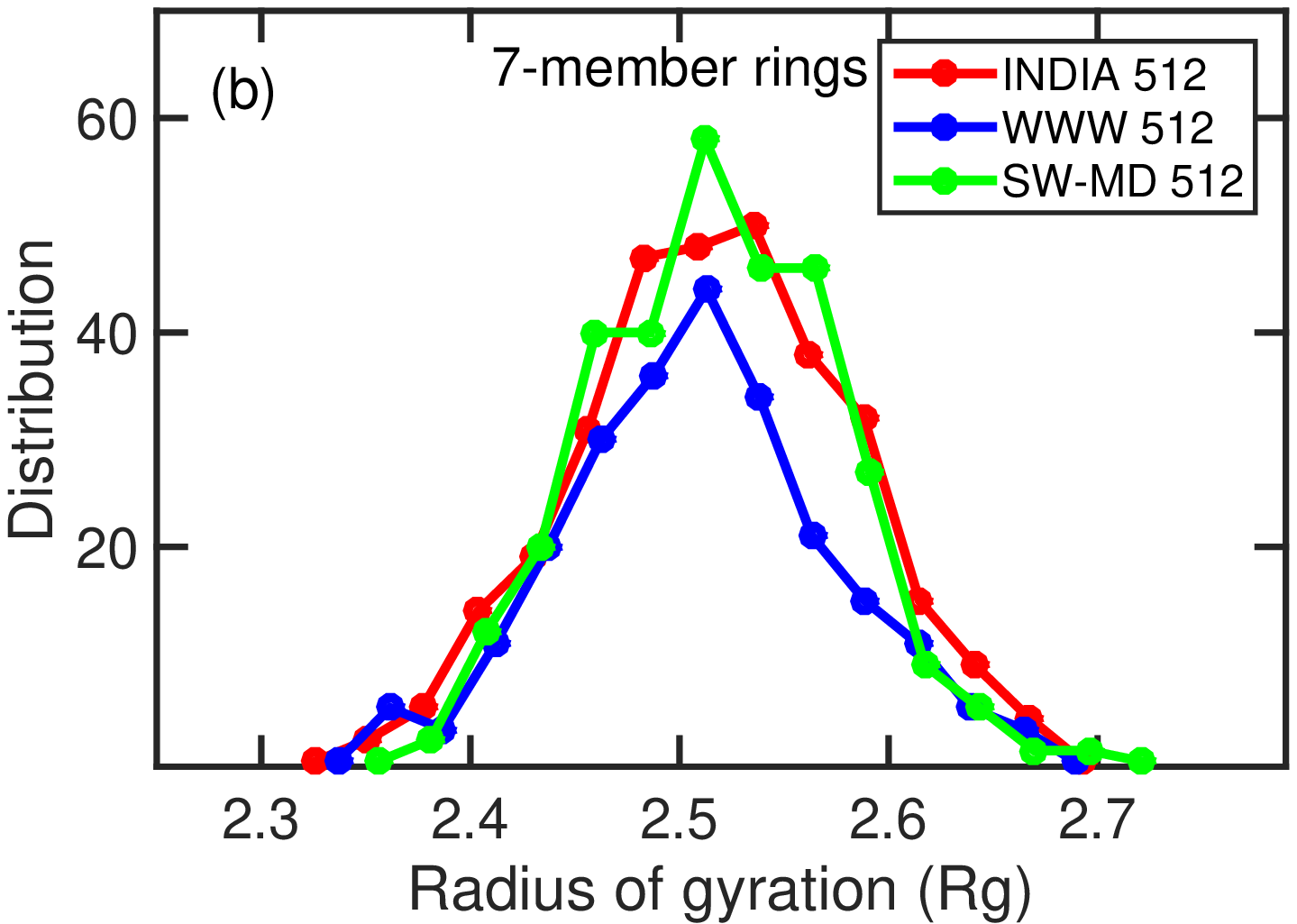}
\caption{\label{fig3}
{\small 
The distribution of the radii of gyration of (a) 6-member rings and (b)
7-member rings for the 512-atom INDIA (red), WWW (blue)  and SW-MD (green) 
models of {\asi}. The $R_g$ value of 6-member rings for ideal {\csi} 
networks is shown by a vertical (green) line in Fig.\,\ref{fig3}(a).
For references to color in this figure legend, see the 
web version of this article. 
}}
\end{figure}

We also examined the topological connectivity and the associated
length scale(s) for various irreducible ring structures present in the amorphous
networks by computing the radius of gyration ($R_g$) of the rings. The
$R_g$ values and the numbers of the high-member rings ($\ge$ 7-member)
can be indicative of the presence of intermediate range order (IRO) in
amorphous networks. The distributions of $R_g$ of 6-member rings and
7-member rings are shown in Fig.~\ref{fig3}(a) and Fig.~\ref{fig3}(b)
for the 512-atom models, respectively. The plot in
Fig.~\ref{fig3}(a) indicates the presence of more 6-member rings in
the SW-MD model compared to the INDIA and WWW models of an identical
size. This can be attributed to the somewhat more ordered nature of
the SW-MD model, which has been obtained from the MD simulations and
thus can be compared with annealed samples of {\asi}. The average
$R_g$ value of 6-member rings in {\csi} is found to be 2.25 {\AA},
which is somewhat lower that the average bond length (of 2.35 {\AA})
in ideal {\csi}, due to the non-planar nature of the hexagonal rings.
Similarly, Fig.\,\ref{fig3}(b) shows the results for 7-member rings 
that are somewhat larger than the corresponding 6-member rings. 

Finally, we addressed the electronic and thermodynamic properties of the
INDIA models. Toward this end, we computed the electronic density of
states and the variation of the specific heat at constant volume with
temperature. Figure~\ref{fig4}(a) shows the density of electronic
states for the 300- and 512-atom INDIA models and compared the results
with the same from the 512-atom WWW model. The INDIA models produced 
a remarkably clean electronic gap, almost identical to the WWW model.  
We studied the specific heat at constant volume 
($C_v$) of the 300- and 512-atom INDIA 
models in the temperature range of 10 K--300 K. The specific heat 
was calculated from the (discrete) vibrational frequencies
($\omega$) obtained from direct diagonalization of the dynamical 
matrices in the harmonic approximation using the 
relation,~\cite{maradudin1971} 
\be
C_v(T) = k_B \sum_{\bm{k},j}
\frac{(\frac{\hbar\omega_j(\bm{k})}{2k_BT})^2}{\sinh^2(\frac{\hbar\omega_j(\bm{k})}{2k_BT})},
\label{E4}
\ee
\noindent 
and compared the resulting values with the experimental 
data obtained by Zink {\etal}~\cite{zink2006} at low 
temperature in the range from 10 K to 300 K.  As mentioned in 
Sect.\,II, the dynamical matrices of the models were obtained 
from employing the density-functional code {\sc Siesta}. 
Given the temperature range studied here, the harmonic 
approximation was found to be adequate in the present 
work. In Eq.\,\ref{E4}, the sum over $\bm{k}$ corresponds 
to the $\Gamma$ point only.  Figure \ref{fig4}(b) shows 
the low-temperature dependence of the specific 
heat ($C_v/T^3$) with temperature from 10 K to 300 K. The 
results suggest that the INDIA models are in 
good agreement with the experimental data for $ T > $ 40 K. 
The inset shows the classical Dulong-Petit limit, $C_v \to 3R$, 
as the temperature approaches to 300 K.

\begin{figure}[t]
\includegraphics[width=0.4\textwidth]{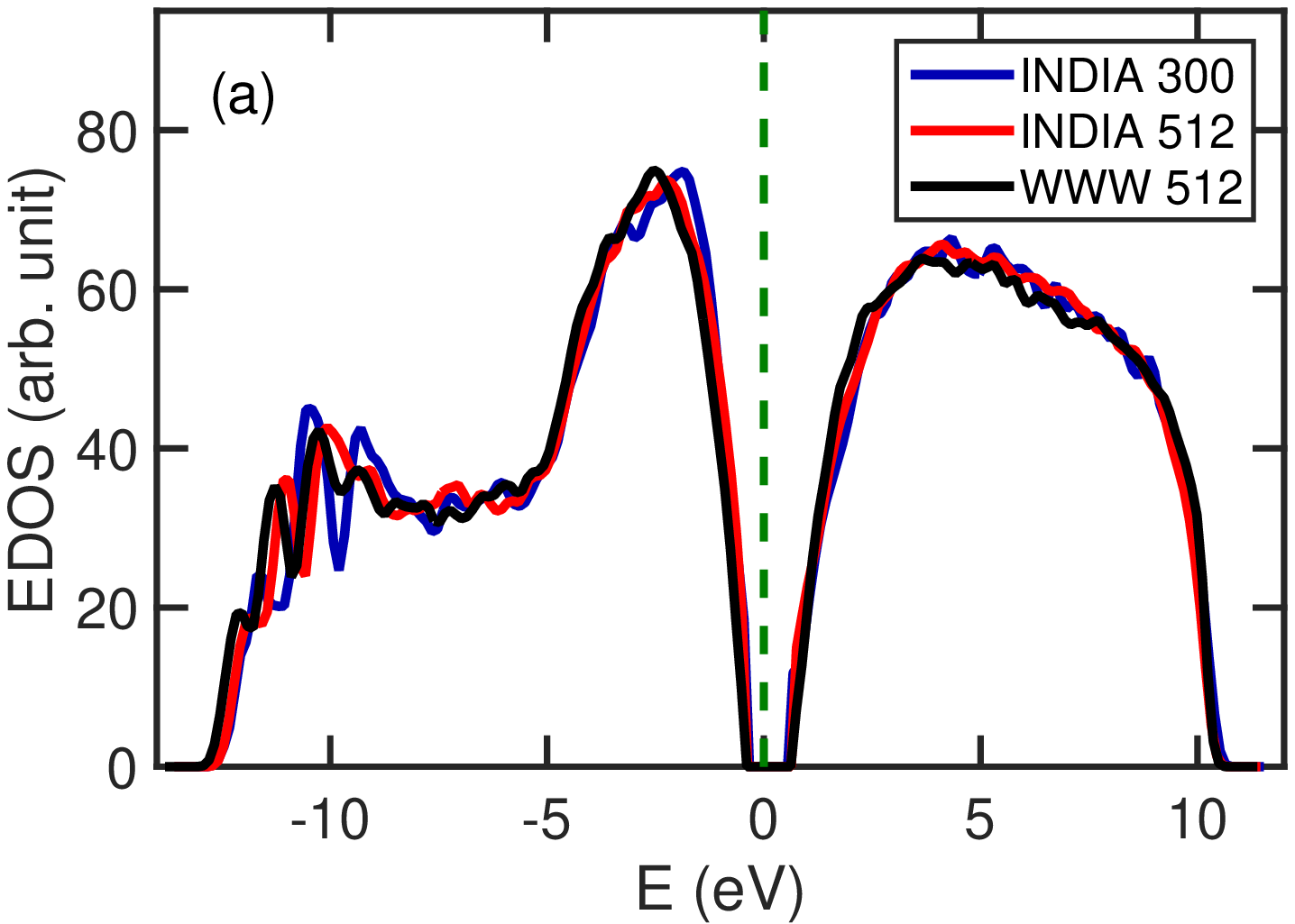}\vspace{2mm}
\includegraphics[width=0.4\textwidth]{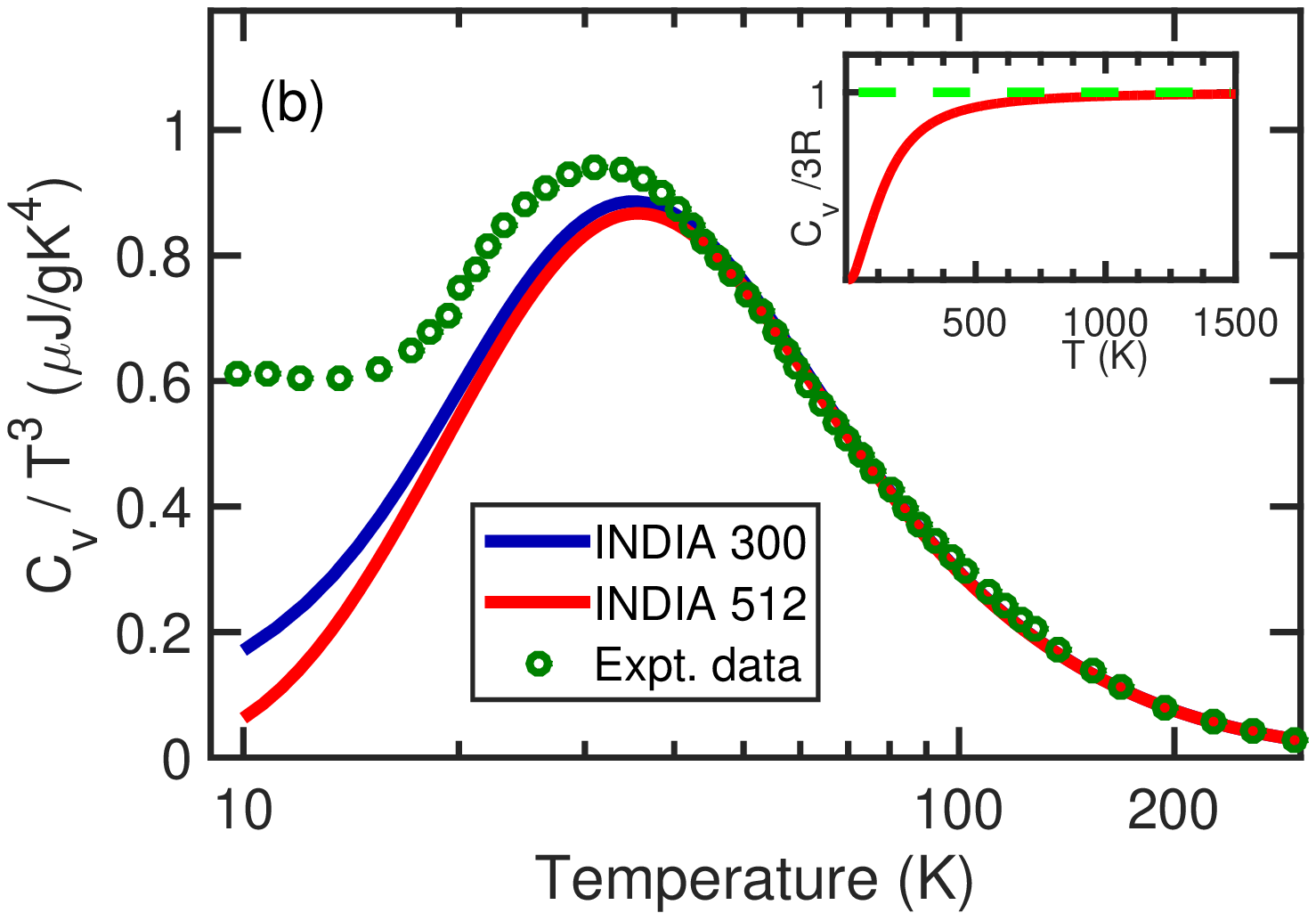}
\caption{\label{fig4}
{\small 
(a) The electronic densities of states of {\asi} from the 
300-atom INDIA (blue), 512-atom INDIA (red) and 512-atom 
WWW (black) models with the Fermi level indicated by a 
vertical dashed line at 0 eV. (b) The low-temperature 
dependence of the specific heat of the 300- and 512-atom 
INDIA models compared with the experimental data (green) from 
Zink {\etal},~\cite{zink2006} reprinted with 
permission from American Physical Society and 
the classical {\textit{Dulong-Petit limit}}. For references 
to color, see the web version of this article.
}}
\end{figure}

\section{Conclusions}
In this paper, we have studied the structural, electronic, vibrational 
and thermodynamic properties of amorphous silicon obtained from 
using an information-driven inverse approach developed in
Ref.\onlinecite{limbu2018}, which simultaneously uses experimental
structure factor and a few structural constraints along with a classical 
total-energy functional. By introducing a subspace optimization
scheme, it was shown that structural constraints can be readily
incorporated to produce high-quality model configurations of {\asi},
which are otherwise very difficult to achieve using the conventional
RMC and hybrid RMC schemes. The resulting data-driven optimization
method produces high-quality (i.e., a few coordination defects and a
narrow bond-angle distribution) CRN models of amorphous silicon of
size up to 1024 atoms. The resulting relaxation-based models have the average
bond angle of 109.2{\degree} and the root-mean-square (RMS) deviation
of 9-11{\degree}, which are within the range of experimental values 
obtained from Raman measurements. The data-driven computational 
approach can produce realistic models of {\asi}, which exhibit
accurate structural properties as produced by 
the WWW method and observed in experiments. The INDIA models show 
not only an excellent agreement with structural properties but 
also able to produce a pristine gap in the electronic density 
of states and the thermodynamic and vibrational properties, compatible 
with experimental measurements.

\section*{Acknowledgments}
The work is partially supported by the US National Science Foundation
(NSF) under Grants No. DMR 1507166 and No. DMR 1507118. PB thanks 
Prof.\,Stephen Elliott (Cambridge, UK) for discussions. 

\input{paper.bbl}

\end{document}

%% file: paper.bbl
%